\documentstyle[epsfig]{l-aa}
\def\ltsima{$\; \buildrel < \over \sim \;$}
\def\simlt{\lower.5ex\hbox{\ltsima}}
\def\gtsima{$\; \buildrel > \over \sim \;$}
\def\simgt{\lower.5ex\hbox{\gtsima}}

\begin{document}
\thesaurus{(03 13.25.2; 11.09.1 1H0419-577; 11.19.1 }

\title{1H0419-577: a ``two-state'' soft X--ray Seyfert galaxy\thanks{Partly based on observations collected at the European Southern Observatory, La Silla, Chile}}

\author{M.Guainazzi\inst{1,2}, A.Comastri\inst{3}, G.M.Stirpe\inst{3}, W.N.Brandt\inst{4}, F.Fiore\inst{5,2}, K.M.Leighly\inst{6}, G.Matt\inst{7}, S.Molendi\inst{8}, E.M.Puchnarewicz\inst{9}, L.Piro\inst{10}, A.Siemiginowska\inst{11}}

\institute{
{Astrophysics Division, Space Science Department of ESA, ESTEC, Postbus 299,
2200 AG Noordwijk, The Netherlands}
\and
{{\it Beppo-SAX} Science Data Center, c/o Nuova Telespazio, Via Corcolle 19 I-00131 Roma, Italy}
\and
{Osservatorio Astronomico di Bologna, Via Zamboni 33, I-40126 Bologna, Italy}
\and
{Department of Astronomy and Astrophysics, Penn State University, 525 Davey Lab, University Park, PA 16802, U.S.A.}
\and
{Osservatorio Astronomico di Roma, Via dell'Osservatorio 5, I-00040 Monteporzio-Catone (RM), Italy}
\and
{Columbia Astrophysical Laboratory, Columbia University, 538 West 120th Street, New York, NY 10027, U.S.A.}
\and
{Dipartimento di Fisica, Universit\`a degli Studi ``Roma 3'', Via della Vasca Navale 84, I-00146 Roma, Italy}
\and
{Istituto di Fisica Cosmica e Tecnologie Relative/C.N.R., Via Bassini 15, I-20133 Milano, Italy}
\and
{Mullard Space Science Laboratory, University College London, Holmbury St.Mary, Dorking, Surrey RH5 6NT, U.K.}
\and
{Istituto di Astrofisica Spaziale - C.N.R., Via Fosso del Cavaliere, I-00133 Roma, Italy}
\and
{Harvard-Smithsonian Center for Astrophysics,60 Garden St., Cambridge, MA 02138, U.S.A.}
}
   
\offprints{M.Guainazzi, mguainaz@astro.estec.esa.nl}

\date{Received  ; accepted }

\maketitle

\markboth{M.Guainazzi et al.}{1H0419-577:...}

\begin{abstract}
In this paper we report on the first simultaneous
optical and X--ray (Beppo-SAX) observations of the radio-quiet
AGN 1H0419-577.
The optical spectrum clearly leads us to classify this source
as a Seyfert 1. The X--ray spectrum is, however, somewhat at odds
with this classification: a simple flat ($\Gamma \sim 1.55$)
and featureless power--law is a good description of the whole
1.8--40~keV spectrum, even if the upper limit to a broad
iron line is not very tight. An analysis
of a still unpublished
ROSAT observation of the same target reveals that the soft X--ray
spectrum has undergone a transition from a steep ($\Gamma = 2.5$)
to a flat ($\Gamma = 1.55$) state, at least in the 0.7--2~keV band.
If this difference is due to a remarkably variable soft excess,
it is unlikely that a single component is responsible for the
optical/UV/soft X--ray spectral distribution. The hypothesis
that the difference is due to a change in the primary X--ray continuum
and its implications for current Comptonization models are discussed.

\keywords{X-ray: galaxies -- Galaxies:Seyfert -- Galaxies:individual:1H0419-577}

\end{abstract}

\section{Introduction}

Soon after it was discovered that the X--ray spectra of Seyfert 1 galaxies
can be described at 0-th order by a simple power--law (Mushotzky 1984),
it became clear that significant deviations exist
from this simple behavior. In particular, a {\it soft excess}
above the extrapolation of the high--energy power--law was observed by all
the main X--ray missions in the 80s: HEAO-1 (Pravdo et al. 1981;
Singh et al. 1985),
EXOSAT (Arnaud et al. 1985; Turner \& Pounds 1989), Einstein
(Kruper et al. 1990; Turner et al. 1990), BBXRT
(Turner et al. 1993).
Optical/UV and soft X--ray emissions are generally well correlated
(Walter \& Fink 1993; Puchnarewicz et al. 1996; Laor et al. 1997)
and this has traditionally supported the idea that they are one and the
same spectral component, probably originating as thermal emission in
an accretion disk (Czerny \& Elvis 1987). However, thin disc
models generally fail to reproduce the observed Spectral Energy
Distribution (SED, Ulrich \& Molendi 1996; Laor et al. 1997).
Moreover, the tight correlation between optical and UV variations in some
Seyfert galaxies (Peterson et al. 1991,
Clavel et al. 1991) argues against an origin as
a thermal emission from accretion discs.
If irradiation of the disc by a high-energy external source is
taken self-consistently into account, the agreement between data
and models improves (Molendi et al. 1992,
Siemiginowska et al. 1995).
There is nowadays formidable evidence that reprocessing of
the primary radiation indeed occurs on $\mu$pc scale around the
nuclei of Seyfert 1s, either from neutral (Pounds
et al. 1990, Piro et al. 1990, Nandra \& Pounds 1994,
Nandra et al. 1997a) or ionized (Piro et al. 1997)
matter.

Given this framework, we discuss in this paper the results of a Beppo-SAX
observation of the radio--quiet AGN 1H0419-577 (1ES 0425-573),
along with the analysis of a still
unpublished pointed PSPC observation, which
was performed four years earlier.
1H0419-577 is classified as a Seyfert 1 by
Brissenden (1989) and as a Seyfert 1.5 by Grupe (1996).
A high-quality optical spectrum was taken the same night
as the Beppo-SAX observation, and the relevant results are
presented in this paper.

1H0419-577 is a moderately distant ($z = 0.103$) object.
It is one of the brightest extragalactic sources in the EUVE
catalog, with flux $F = 3.6 \times 10^{-12}$~erg~cm$^{-2}$~s$^{-1}$
in the 58--174~\AA band (Marshall et al. 1995, Fruscione
1996). It is also included in the Seyfert soft X--ray selected
sample of Grupe et al. (1998, RXJ0426-57).
No significant radio emission is reported, the upper limit
on the 8.4~GHz flux being $3.6$~mJy (Brissenden 1989). Despite the
fact that it is a relatively bright source in X--ray (it was
detected by HEAO A-2, Mushotzky 1995, private communication),
its X--ray properties have not been well studied yet.

The paper is organized as follows: in Sect.~2 the optical data are
presented. Reduction and analysis of the Beppo-SAX data are presented in
Sect.~3 and Sect.~4 respectively. In Sect.~5 the PSPC data are analyzed
and the relevant outcomes are compared with Beppo-SAX ones.
The discussion of the main results is given in Sect.~6.
Sect.~7 summarizes the main conclusions.
$H_0 = 75$~km~s$^{-1}$~Mpc$^{-1}$ is
assumed throughout the paper. Preliminary results have been published in
Guainazzi et al. (1998).

\section{The optical spectrum}

Optical spectra of 1H0419$-$577 were obtained at the ESO 1.52m telescope on 1996
October 1 and~2, using the Boller \& Chivens spectrograph with a 127mm camera
and a Loral/Lesser thinned CCD with $2048\times2048$ pixels.  The pixel size is
$15\mu\hbox{m}$, and the projected scale on the detector
0.82~arcsec~pixel$^{-1}$.  The grating used has 600~grooves~mm$^{-1}$.  The
spectra were obtained through a 2-arcsec wide slit, at a resolution of 4.6~\AA.
Both nights were photometric. Two integrations of 1800 seconds each were 
obtained.

\begin{figure}
\epsfig{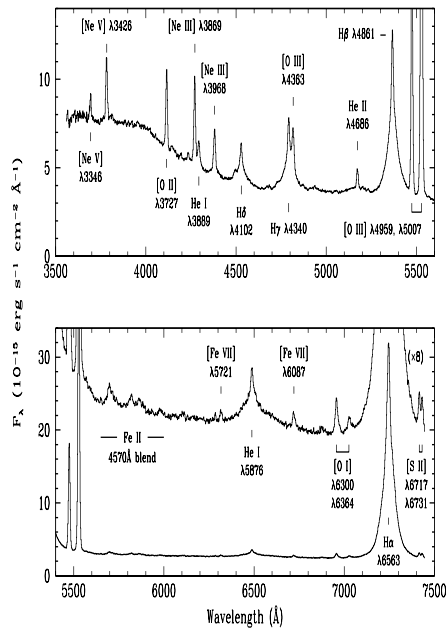}
\caption{The optical spectrum of 1H0419$-$577 obtained in 1996 October. 
The spectrum has been displayed in two parts for clarity, and the red part
(bottom panel) is also shown magnified by a factor 8 to enhance the weaker
lines.  The main optical lines are indicated on the plot. 
The wavelength scale is in the rest system of 
the observer (z=0.103).}
\label{fig9}
\end{figure}

Standard techniques were used to reduce the spectra, using the NOAO IRAF
package.  In particular, the spectrophotometric calibration of the spectra was
perfected through their comparison with, and broad-band normalized to, short
exposures of 1H0419$-$577 obtained during the same nights through an 8~arcsec 
slit.
This allowed us to correct the narrow-slit spectra for light losses and
differential refraction.  The two spectra were averaged and
yielded the final spectrum shown in Fig.\ref{fig9}.  Because the calibration curves
of the two nights and the individual spectra coincide within $<5$\%, we
consider the spectrophotometric quality of the final spectrum to be of this 
order.

The spectrum displays characteristics typical of Seyfert 1 galaxies, with strong
broad permitted lines and narrow forbidden lines. Table~\ref{tab5} lists the fluxes,
\begin{table*}
\begin{footnotesize}
\begin{center}
\begin{tabular}{lccc} \hline
Line                       & Flux            & FWHM     & EW \\
                           & (erg/s/cm$^2$)    & (km/s)   & (\AA) \\ \hline \hline
$[$Ne V$] \lambda$3346 &                            $1.1\times10^{-14}$ &         620   &   2 \\
$[$Ne V$] \lambda$3426 &                           $3.6  \times10^{-14}$ &        760  &    5 \\
$[$O II$] \lambda$3727 &                           $4.8\times10^{-14}$ &         750   &   8  \\
$[$Ne III$] \lambda$3869 &                         $5.7\times10^{-14}$ &    770   &  11 \\
He I $\lambda$3889 + H$\zeta$ &                  $2.4\times10^{-14}$ &  1090    &  5 \\
$[$Ne III$] \lambda$3968 + H$\eta$ &               $4.7\times10^{-14}$    &      950  &   10 \\
H$\delta$ + $[$S II$] \lambda$4068,$\lambda$4076 &  $7.7 \times10^{-14}$   &    1500   &  18 \\
H$\gamma$ broad &                               $1.9   \times10^{-13}$   &    3470  &   51 \\
H$\gamma$ narrow &                              $2.0    \times10^{-14}$  &   560    &  5 \\
$[$O III$] \lambda$4363  &                          $2.3  \times10^{-14}$  &        670   &   6 \\
He II $\lambda$4686 broad  &                     $2.4   \times10^{-14}$  &      3720   &   7 \\
He II $\lambda$4686 narrow &                    $1.2 \times10^{-14}$ &          600  &    4 \\
H$\beta$ broad &                                $4.7    \times10^{-13}$ &       3460  &  146 \\
H$\beta$ narrow &                               $5.0  \times10^{-14}$ &     560   &  16 \\
$[$O III$] \lambda$4959 &                           $1.7 \times10^{-13}$ &        560  &   53 \\
$[$O III$] \lambda$5007 &                          $5.0 \times10^{-13}$ &       560  &  160 \\
He I $\lambda$5876 broad &                      $8.4    \times10^{-14}$ &      4650 &    32 \\
He I $\lambda$5876 narrow &                     $5.0   \times10^{-15}$ &       560  &    2 \\
$[$Fe VII$] \lambda$6087 &                         $4.0 \times10^{-15}$ &  730   &   2 \\
$[$O I$] \lambda$6300 &                            $9.7      \times10^{-15}$ &    720 &     4 \\
$[$O I$] \lambda$6364 &                            $3.4 \times10^{-15}$ &     720   &   1 \\
H$\alpha$ broad &                               $1.6 \times10^{-12}$ &         2660  &  704 \\
H$\alpha$ narrow &                              $1.1  \times10^{-13}$ &       560  &   50 \\
$[$S II$] \lambda$6717 &                           $6.6    \times10^{-15}$ &      460  &    3 \\
$[$S II$] \lambda$6731 &                           $6.7 \times10^{-15}$ &         460  &    3 \\ \hline \hline
\end{tabular}
\end{center}
\end{footnotesize}
\caption{Integrated fluxes, FWHM, and equivalent widths of the main optical emission lines}
\label{tab5}
\end{table*}
equivalent widths, and FWHM of the main lines.  In order to obtain these
measurements a 1st order pseudo-continuum was fitted to the local minima
bracketing each line.  In most permitted lines the broad component was isolated
by modeling the narrow line contribution with a suitably scaled profile of the
[O~III]$\lambda$5007 line.  Each scaling factor was determined on the basis of
the smoothest residual broad profile.  The corresponding line fluxes in
Table~\ref{tab5} are thus simply the [O~III]$\lambda$5007 flux multiplied by the respective
scaling factors.  In addition, the contaminating lines were subtracted from
H$\alpha$ and H$\beta$ using a scaled [O~III]$\lambda$5007 profile for the
narrow lines ([O~III]$\lambda$4959, [N~II]$\lambda$6548 and $\lambda$6584)
and a scaled broad-H$\alpha$ profile for the broad lines (He~II~$\lambda$4686
and Fe~II~$\lambda$4924).  The Fe~II lines are weaker than average for a Seyfert
1:  the ratio between the integrated fluxes of the 4570~\AA\ (rest wavelength)
blend and of H$\beta$ is lower than 0.1 (cfr. Joly 1988).

The spectrum is in broad agreement with the one presented
by Brissenden (1989) and confirms unambiguously the classification
of 1H0419-577 as a Seyfert 1.

\section{Beppo--SAX data reduction}

Beppo-SAX (Boella et al. 1997a)
is a major mission of the Agenzia Spaziale Italiana (ASI),
with participation of the Netherlands Agency for Aerospace Programs (NIVR);
it aims to study celestial
sources in the wide X--ray band 0.1--300~keV. The scientific payload
is constituted of a pair of gas scintillator proportional counters
with imaging capabilities
(Low Energy Concentrator Spectrometer, LECS, Parmar et al. 1997,
sensitive in the 0.1--10~keV band;
Medium Energy Concentrator Spectrometer, MECS, Boella et al. 1997b,
sensitive in the 1.8--10.5~keV band), and a pair of collimated
high--energy instruments (High Pressure Gas Scintillating Proportional
Counter, HPGSPC, Manzo et al. 1997, sensitive in the
7--60~keV band; Phoswitch Detector System,
PDS, Frontera et al. 1997, sensitive in the 13--200~keV band).
The HPGSPC is more oriented to high-resolution spectroscopy of
bright sources, while the PDS possesses an unprecedented sensitivity in its
energy range (Guainazzi \& Matteuzzi 1997).

Beppo-SAX observed 1H0419-577 on 1996 September
30, from 06:24:00 UT to 16:40:50 UT. The LECS was switched off during
the whole observation due to technical problems.
The other instruments were operating in direct modes, which provide full
information on the arrival time, energy, burst length/rise time and - if
available - position of each photon. Data have been reduced with the
{\sc Saxdas} package (version 1.1) in the {\sc Ftools 3.5} environment. Standard
selection criteria on satellite aspect quantities were applied to
avoid Earth occultation (angle between the pointing direction and the
Earth's limb $\ge 5^{\circ}$); South Atlantic Geomagnetic Anomaly (SAGA)
passages; and particle contamination (momentum associated to the
Geomagnetic cut-off Rigidity $\ge 6$~GeV/c).
In the MECS image, the source is nearly on-axis (offset angle $\simlt 1'$). 
MECS scientific products
have been extracted from a circular area 3' radius around the apparent
centroid of the source. Background spectra have been extracted from blank
sky fields, using the same extraction region in detector coordinates as
the source. Appropriate response matrices have been created, using the
calibration files publicly available since 1997 January 31, which include
the results of on-flight calibration. Total net MECS exposure time is
$T \simeq 22600$~s and the corresponding count rate is
$0.171 \pm 0.003$~s$^{-1}$ in the whole energy band.
The spectrum has been rebinned in order to have at least 20
counts per channel, to ensure the applicability of $\chi^2$ statistics.
The source is not detected in the HPGSPC.
PDS data have been further screened
by eliminating 5 minute intervals after any SAGA passage,
in order to avoid gain instabilities due to
recovery to the nominal values after
instrument switch off. The PDS was working in rocking mode, each
collimator pointing alternatively every 96 seconds to
the source and a field $\pm 3.5^{\circ}$
aside.
Short-lived ($\tau \simlt 1$~s) spikes, due to particle ``bursts'' along
the orbit, were also removed, according to the following procedure:
bins in the 2-seconds binned light curves of each detector, whose count rate
exceeded the mean by
5 standard deviations, have been excluded from the scientific
product accumulation.
Net exposure time in the PDS results
$\sim$ one half of the MECS ($\simeq 10700$~s). In the 13--36~keV band
the net background-subtracted count rate is $0.129\pm0.052$~s$^{-1}$;
a typical systematic uncertainty
of $\sim 0.02$~s$^{-1}$ in the whole 13--200~keV PDS band
has been taken into account hereinafter.
(Guainazzi \& Matteuzzi 1997). The 13--36~keV
detection is significant at the
5~$\sigma$ level.

Data analysis has been performed using the {\sc Xanadu} package. Hereafter
energies are quoted in the source rest frame and errors are 90\% confidence
level for one interesting parameter ({\it i.e.} $\Delta \chi^2 = 2.71$),
unless otherwise specified.

\section{Beppo--SAX data analysis}

\subsection{Timing analysis}

\begin{figure}
\epsfig{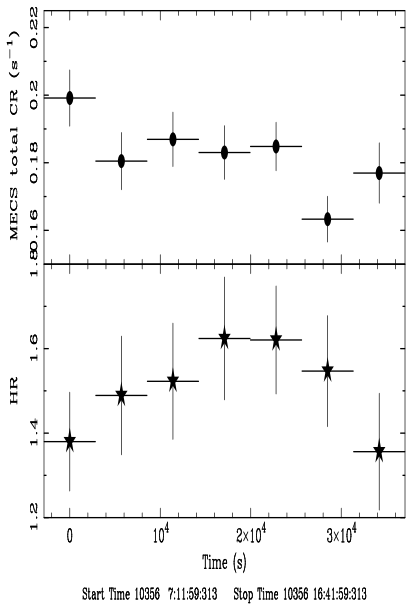}
\caption{Broadband ({\it i.e.} 1.8-10.0~keV) MECS light curve ({\it upper panel})
and ratio ({\it lower panel}) between the 3--10.5 and 1.8--3.0~keV bands count rates.
Binning time
is $\Delta t = 5700$~s, corresponding approximately to one Beppo-SAX
orbit.}
\label{fig1}
\end{figure}

In Fig.~\ref{fig1} the full energy band MECS light curve of the Beppo-SAX
observation is shown. It displays a smooth decreasing trend of $\simeq 15\%$
during the $\sim 3.5 \times 10^4$~s of elapsed time.
The hardness ratio between the count rates in the
1.8--3.0 and 3.0--10.5~keV bands
(see Sect.~4) shows a regular trend, increasing in the first
$2 \times 10^4$~s and decreasing thereafter.
However, the $\chi^2$ for constant hypothesis is 4 over 6
degrees of freedom only.
We extracted spectra during the time intervals when the HR is
higher or lower than the mean value. The difference in the
spectral indices of a simple power--law model between the
two states is $\simeq 0.06$, comparable with the statistical
uncertainties.
We will therefore focus in the following
on the time-averaged spectral behavior to achieve the maximum S/N
ratio.

\subsection{Spectral analysis}

\begin{figure}
\hspace{0.5cm}
\epsfig{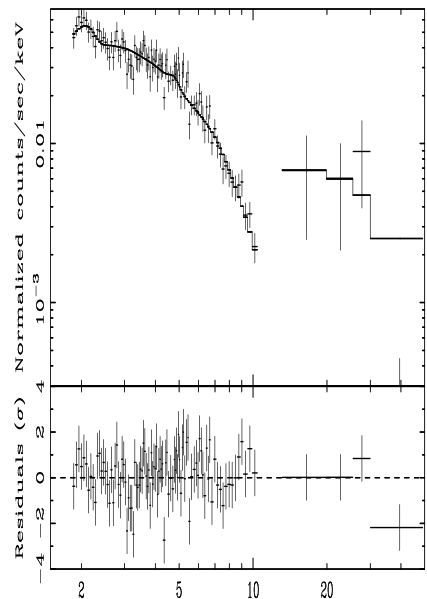}
\caption{Spectrum ({\it upper panel}) and residuals in units of standard
deviations ({\it lower panel}) when a simple power-law
model with photoelectric absorption by cold matter is applied to the MECS
and PDS data simultaneously. Each of the PDS data points has a S/N $> 1.5$,
each of the MECS data points a S/N $> 5$.}
\label{fig2}
\end{figure}

A simple power-law with photoelectric absorption by cold matter
is a rather good representation of the MECS spectrum
(see Fig.~\ref{fig2}).
If the absorbing column density is left free to vary,
$N_H < 2.3 \times 10^{21}$~cm$^{-2}$.
$N_H$ has been therefore constrained to be not lower
than the
Galactic value along the 1H0419-577 line of sight ($N_{H_{Gal}} =
2 \times 10^{20}$~cm$^{-2}$, Dickey \& Lockman 1990) and 
is always consistent with its
minimum allowed value. The $\chi^2$ is formally acceptable
($\chi^2 = 114/124$ d.o.f.).
The spectral index is rather flat ($\Gamma_{3-10 keV}
= 1.51^{+0.09}_{-0.12}$) if compared with the mean value for
the Seyfert 1 Galaxies
($\Gamma_{3-10 keV} = 1.91 \pm 0.07$, Nandra et al. 1997a)
observed by ASCA.
The unabsorbed flux in the 2-10~keV band is $F = (1.1\pm0.2) \times
10^{-11}$~erg~s$^{-1}$~cm$^{-2}$ ($\sim 0.5$~mCrab),
corresponding to a luminosity
$L \simeq 5.3 \times 10^{44}$~erg~s$^{-1}$, which ranks 1H0419-577
among the most luminous Seyfert 1s in X--rays. The normalization at 1~keV is
$(2.32\pm0.12) \times 10^{-3}$~photons~cm$^{-2}$~s$^{-1}$.

If we superimpose the PDS points to the power-law best fit, they
lie well on the power-law extrapolation ($\chi^2 = 131/153$ d.o.f.), provided
the usual relative normalization
constant $C_{PDS/MECS}=0.80$ between MECS and PDS flux at 1 keV is assumed
(Cusumano et al. 1998). The results are not affected by the residual
$\pm 10\%$ uncertainty on $C_{PDS/MECS}$.

Although we have not found clear evidence of any deviation from the
simple power-law behavior, we have searched for narrow line
emission features and/or changes in the continuum curvature in the broadband
spectrum. If a
broken power-law is used instead of a simple power-law, the $\chi^2$ is
only marginally better [F-test ($F$) equal to 2.6, significant at
only 96.0\%]. We have also tried a more physical
double power-law model, which results in a (basically
unconstrained) very steep soft component and a
$\Delta \chi^2 = 4.1$.
Although the improvement in the quality of the fit is not
statistically negligible (99.7\% significance level),
an inspection by eye of the residuals (see Fig.~\ref{fig2})
reveals that such a solution tends to account for the behavior of the
lowest energy channels 
in the MECS bandpass. The deviation expressed
in data/model ratio is there $\sim 10\%$, {\it i.e.} of the same order of the
calibration uncertainties; moreover, the feature below 2~keV
is {\it not}
the widest regular one in the residuals spectrum.
We consider therefore such evidence as scarcely conclusive
and will search for more robust and calibration-independent tests
for it.
The attempts at modeling the slight concavity
of the spectrum at low energy either with thermal or partial
covering models were even less
successful. Table~\ref{tab1} reports a summary of the fit results.
\begin{table*}
\begin{footnotesize}
\begin{center}
\begin{tabular}{lccccc} \hline \hline
\multicolumn{6}{c}{Panel a: power-laws continuum modeling} \\ \hline
Model & $\Gamma$ & $\Gamma_{soft}$ & $E_{break}$ & $N_{soft}/N_{hard}$ & $\chi^
2$/d.o.f. \\
& & & (keV) & & \\ \hline
PO & $1.61\pm0.06$ & ... & ... & ... &130.6/153 \\
BKNPO & $1.56\pm0.10$ & $1.8\pm0.4$ & $3.0\pm1.0$ & ... & 128.0/151 \\
PO+PO & $1.55\pm0.09$ & $7^{\ddag}$ & ... & 3.7$^{\ddag}$ & 126.5/150 \\ \hline
\multicolumn{6}{c}{Panel b: iron line emission} \\ \hline
Model & $\Gamma$ & $E$ & $\sigma$ & $EW$ & $\chi^2$/d.o.f. \\
& & (keV) & (keV) & (eV) & \\ \hline
PO+GA & $1.64^{+0.09}_{-0.08}$ & $6.2^{\ddag}$ & $0.7^{\ddag}$ & $180^{\ddag}$ & 128.1/151 \\
Narrow neutral line & $1.61^{+0.07}_{-0.06}$ & 6.4$^{\dag}$ & 0$^{\dag}$ & $17^{+83}_{-17}$ & 130.6/152 \\
Broad neutral line & $1.65\pm0.08$ &  6.4$^{\dag}$ & $0.43^{\dag}$ & $< 250$ & 128.4/151 \\ 
Narrow ionized line & $1.63^{+0.06}_{-0.07}$ & 6.7$^{\dag}$ & 0$^{\dag}$ & $70^{+90}_{-70}$ & 129.1/152 \\
Broad ionized line & $1.63^{+0.10}_{-0.07}$ & 6.7$^{\dag}$ & 0.43$^{\dag}$ & $<260$ & 129.1/151 \\ \hline \hline
\end{tabular}
\\
\noindent
$^{\ddag}$unconstrained \\
$^{\ddag}$fixed
\end{center}
\caption
{Results of the simultaneous fits of MECS and PDS data. A photoelectric absorption
from cold
matter with $N_H=N_{H_{Gal}}=2 \times 10^{20}$~cm$^{-2}$ was added to all the
quoted models.
PO=power-law, BKNPO=broken power-law, GA=Gaussian line.}
\label{tab1}
\end{footnotesize}
\end{table*}

Broad K$_{\alpha}$ fluorescent lines from neutral or low-ionized
iron have been detected in almost all the Seyfert 1s observed by Ginga
(Nandra \& Pounds 1994) and ASCA (Nandra et al. 1997a) so far.
Adding a narrow
({\it i.e.} $\sigma = 0$) Gaussian emission
profile to the simple power-law model improves only marginally
the fit ($F=1.9$, significant at 84.6\%).
If the centroid energy of the line is frozen
at 6.4~keV (neutral iron) or 6.7~keV (He-like iron),
the equivalent widths (EW) are $EW_{6.4~keV}= 17^{+83}_{-17}$~eV and
$EW_{6.7~keV}=70^{+90}_{-70}$~eV respectively.
If the widths of the lines are allowed to vary as free
parameters, the decrease of the $\chi^2$ in comparison to the
narrow--line case is always negligible ($\Delta \chi^2 < 1$).
When the widths of the lines are held fixed to the mean
in Nandra et al. sample (430~eV, 1997a), the upper limits
on the EW are $\simeq 250$~eV.

Flattened spectra can be produced if a
Compton reflection component is superimposed on a steeper
intrinsic continuum.
This hypothesis is supported in Seyfert galaxies by
the flattening of the continuum shape at $E \simgt 10 \
keV$ observed by Ginga (Pounds et al. 1990, Piro
et al. 1990) and by the detection of broad fluorescent iron
lines
(Tanaka et al. 1995; Nandra et al. 1997a),
which are considered to be signatures of reprocessing of the
nuclear X--rays by optically thick matter, possibly in the
form of a rotating accretion disk around the central black hole. We have
tested such an hypothesis with the model {\verb!pexrav!} in {\sc Xspec},
leaving as free parameters only
the intrinsic spectral index and the relative normalization $R$ between
the direct and the reflected component. The other parameters are basically
unconstrained and we have therefore fixed the cut-off energy of the direct
spectrum at $E_{cut-off} = 100$~keV (Gondek et al. 1996)
and the angle between the disk axis and the line-of-sight at
$30^{\circ}$ (Nandra et al.
1997a).
The improvement of the fit is significant at 98.4\% ($F=5.9$),
and the intrinsic photon index turns out to be even steeper than typical
values observed in Seyfert galaxies ($\Gamma \simeq 2.15$). The
nominal best--fit values correspond to an unplausibly high
amount of reflection
($R = 10$) but the spectral parameters are basically unconstrained.
The upper limits on the EW of a narrow iron line added to such a
continuum
are $EW_{6.4~keV} < 40$~eV and $EW_{6.7~keV} < 90$~eV
in the ``neutral'' and ``ionized''
cases, respectively.  Leaving the widths of the lines free
results in no ({\it i.e.} $\Delta \chi^2
= 0$) further improvement.
If we fix the intrinsic spectral index to the average value found
by Nandra et al. (1997a), $R = 3.8\pm0.9$, whereas the EW
of a narrow (broad) neutral fluorescent line is $<110$ (200)~eV.
It is therefore unlikely that the line originates in the same relativistic
X--ray illuminated disc which could be advocated as the responsible for the
huge continuum reflection component, unless the disc is substantially ionized.
We consider hereafter a simple power-law with photon
index $\Gamma \simeq 1.55$ the best modeling of the spectral shape
in the whole 1.8--40~keV band.

\section{ROSAT PSPC observation}

\subsection{ROSAT data analysis}

In order to cope with the lack of information at energies lower than
$\simeq 2 \ keV$, we have analyzed a still
unpublished ROSAT PSPC observation of the same target. 1H0419-577
was observed by ROSAT
on 1992 April 7, for a total exposure time of 4094 seconds.
Source spectra have been extracted from a circular region of 2'45''
radius around the centroid of the source, while background spectra have been
extracted from an annulus of internal and external radii 6'20''
and 8'30'' respectively. The radii have been chosen in order to avoid six
serendipitous sources that have been detected in the field of view.
To perform spectral fitting,
an appropriate PSPC-B
redistribution matrix
for the observation date ({\verb!pspcb_gain2.rmf!}) has been
used and the effective area calculated
with the software tool {\verb!pcarf!}
included in {\sc Ftools 3.6}. Spectral PI channels lower than 11
($E \simlt 0.15$~keV) and 
higher than 32 ($E \simgt 2.1$~keV) in the rebinned spectrum
have been excluded from the
spectral analysis. The resulting net count rate is $5.72\pm0.04$~s$^{-1}$;
background contributes less than 1/30 for each spectral channel.

\begin{figure}
\epsfig{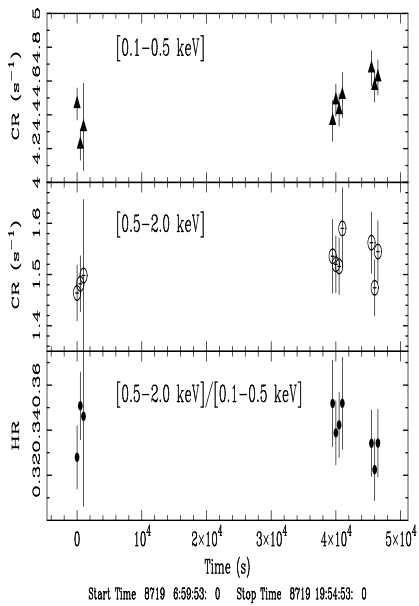}
\caption{PSPC 1H0419-577 light curves in the 0.1--0.5~keV ({\it upper panel})
and 0.5--2.0~keV ({\it central panel}) bands, and their hardness ratio
({\it lower panel}). Binning time is $\Delta t = 500$~s.
The panels are displayed in the same relative ordinate scale to
allow direct comparison of the dynamical ranges.}
\label{fig8}
\end{figure}

The broadband 0.1--2.4~keV light curve shows a slight ({\it i.e.} $\sim 5\%$)
tendency to increase with time. We have investigated whether such a
trend could be associated to spectral variability by studying
the light curves in the 0.1--0.5~keV and 0.5--2.0~keV energy bands
(see Fig.~\ref{fig8}). The suggestion of a wider
dynamical range of the soft light curve variation is only marginal.
In the following we will therefore focus on the
average spectral behavior only.

\begin{figure}
\epsfig{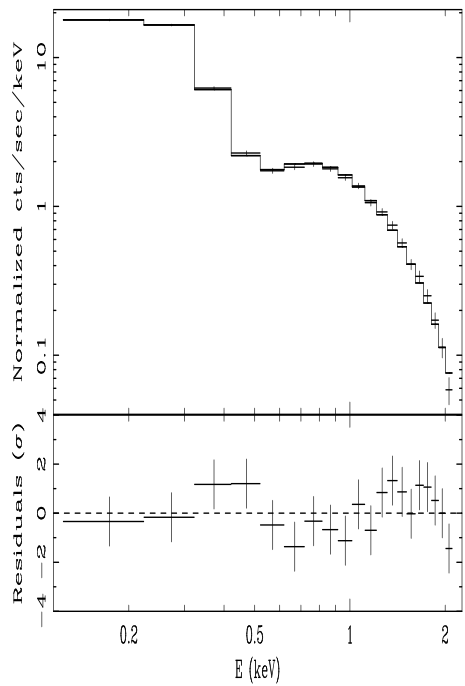}
\caption{Spectrum ({\it upper panel}) and residuals
({\it lower panel}) in units of standard deviations when a simple
absorbed power-law model is applied to the ROSAT PSPC data of 1H0419-577.}
\label{fig3}
\end{figure}

A simple power--law with photoelectric absorption yields a formally
acceptable fit ($\chi^2_r = 0.92$).
$\Gamma = 2.67\pm0.05$ and the 1~keV normalization is $1.03\pm0.03 \times
10^{-2}$~photons~cm$^{-2}$~s$^{-1}$.
The residuals show however
a regular wavy trend (see Fig.~\ref{fig3}).
This trend cannot be removed with a linear adjustment
of the instrumental gain, as achievable with the command
{\verb!gain!} in {\sc Xspec}.
We checked therefore the statistical significance
of spectral complexity. An absorption edge (with energy broadly consistent
with the photoionization threshold energy for {\sc O vii}), a thermal optically
thin or thick emission or a break in the power--law shape yield a
comparable improvement of the $\chi^2$ ($\Delta \chi^2 = 9.7-12.3$
for two further degrees of freedom, corresponding to a confidence level
$\simgt$ 99.97-99.99\%). There is therefore a suggestion of soft X--ray
spectral complexity, with an ultrasoft component emerging below
$\simeq 0.5 \ keV$,
or, alternatively, an absorption edge due to highly ionized interposed matter.
\begin{table*}
\begin{footnotesize}
\begin{center}
\begin{tabular}{lccccc} \hline \hline
\multicolumn{6}{l}{$N_H$ free} \\
Model & $N_H$ & $\Gamma_{soft}$ & $\Gamma_{ultrasoft}$ or $\tau$ & $E$ or $kT$ & $\chi^2/$d.o.f. \\
& ($10^{20} \ cm^{-2}$) & & & (eV)& \\ \hline
PO & $1.42\pm0.11$ & $2.71\pm^{+0.05}_{-0.06}$ & ... & ... & 20.1/17 \\
PO+BB & $1.3^{+0.3}_{-0.2}$ & $2.51^{+0.14}_{-0.16}$ & ... & $73^{+12}_{-29}$ & 9.8/15 \\
PO+BREMS & $1.5^{+0.3}_{-0.2}$ & $2.51^{+0.15}_{-0.27}$ & ... & $150^{+60}_{-90}$ & 10.4/15 \\
BKNPO & $1.64^{+0.19}_{-0.17}$ & $2.54^{+0.12}_{-0.18}$ & $2.88^{+0.22}_{-0.12}$ & $700^{+300}_{-200}$ & 9.9/15 \\
PO+PO	& $1.8^{+1.2}_{-0.4}$ & $2.4^{+4.8}_{-2.4}$ & $3.9^{+3.4}_{-3.9}$  & ... & 11.1/15 \\
ED*PO & $1.53^{+0.15}_{-0.13}$ & $2.71\pm0.07$ & $0.29\pm0.15$ & $0.67\pm0.10$ & 8.8/15 \\ \hline \hline
\end{tabular}
\end{center}
\end{footnotesize}
\caption{Best-fit results of the ROSAT PSPC spectral analysis.
PO=power-law,
ED=absorption edge, BKNPO=broken power-law,
BB=blackbody, BREMS=thermal bremsstrahlung.}
\label{tab3}
\end{table*}
In some fits, the best--fit $N_H$ value is
significantly lower than the Galactic $N_H$ in the direction of
1H0419-577 ($2.0 \times 10^{20}$~cm$^{-2}$), however not inconsistent with it if we take
into account a conservative estimate of the uncertainties on the
nominal Galactic values of 20-30\% ({\it i.e.}: $5 \times 10^{19}$~cm$^{-2}$,
Dickey \& Lockman 1990). 
An investigation of the spectral complexity in the soft X--ray spectrum
would require much better spectroscopic quality and/or simultaneous
monitoring of the soft and hard X--ray.
However, we stress that
the
ROSAT/PSPC spectrum {\it is always much steeper than the MECS
one}, regardless of the specific model adopted. $\Delta \Gamma
\equiv \Gamma_{PSPC} - \Gamma_{MECS} = \Gamma_{soft} - \Gamma_{hard}$
(see Sect.~4) ranges from 0.8 to 1.3.

We investigated also whether we could get comparably good fits if an
underlying MECS-like power--law is assumed.
We than fitted the PSPC spectrum with two-components models,
where one of the components is a power--law with photon
index held fixed at 1.55 and free normalization. The only case for
which an acceptable $\chi^2$ is obtained is the double power--law
model, which yields: $\chi^2 = 11.8/16$~d.o.f., $\Gamma_{soft} = 3.01 /pm 0.18$
and $N_H = (1.7 \pm 0.2) \times 10^{20}$~cm$^{-2}$.
It is better than the single power--law model at the 99.6\%
level of confidence (cfr. first row in Table~\ref{tab3}).
The ratio between the ``hard'' power--law
normalization as measured by the MECS and PSPC
is $1.1 \pm 0.5$.
On the other hand, adding a MECS-like power--law to the other models listed
in Table~\ref{tab3} yields a negligible ({\it i.e.}:
$\Delta \chi^2 < 1$) improvement of the quality of the fit.
In particular, a double blackbody + power--law
model with photon index held fixed to the high--energy
measured value
- which have often been used to approximate the
superposition of a high--energy non--thermal continuum and
a disc emission - is not a viable model ($\chi^2 = 76.8/14$ d.o.f.).

\subsection{ROSAT PSPC/Beppo-SAX MECS comparison}

Beppo-SAX observed 1H0419-577 about four years later than ROSAT.
The source was caught with very different spectral properties;
namely
the ROSAT PSPC 1992 spectrum is much steeper than the 1996 Beppo-SAX MECS one.

Recently, several authors have pointed out that the spectral indices as
measured by the PSPC are systematically steeper than measured by other
missions with a sensitivity bandpass extending at higher energies.
The amount of such a difference has been estimated
between 0.2 (Turner 1993, Fiore et al. 1994) and 0.4
(Iwasawa et al. 1998). It cannot therefore
account completely for the observed difference. Moreover,
If we fit the PSPC data with a power--law of index $\Gamma = 2.0$
and a blackbody component, the $\chi^2$ turns out to be unacceptably
high (439/16 d.o.f.). However, a double power--law model
with one of the photon index held fixed to 2.0
yields $\chi^2 = 11.4$/16~d.o.f., with $N_H = (1.7 \pm 0.2) \times
10^{20}$~cm$^{-2}$ and $\Gamma_{soft} = 3.2 \pm 0.3$.

The energy interval where the two PSPC and the MECS  are well
calibrated and overlap (1.8--2.0~keV) is too narrow to allow a direct
comparison of the spectra via standard fitting techniques.

\begin{figure}
\epsfig{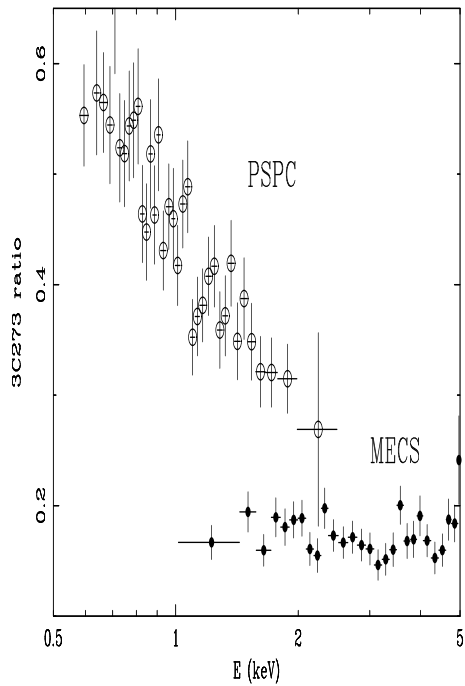}
\caption{Ratio of the PSPC ({\it open circles}) and MECS ({\it filled
circles} spectra with the relevant 3C273 spectra.
Each point corresponds to a S/N $> 10$}
\label{fig6}
\end{figure}

In order to partly overcome such problems, we studied the ratio between
the PSPC and MECS 1H0419-577 spectra and the corresponding spectra of the
featureless radio-loud quasar 3C273, whose X--ray spectrum is known
to be well represented by a simple power-law in the whole 1--200~keV
range (Grandi et al. 1997). The usage of the ratio washes out
in principle any problem
regarding instrumental calibration and the - until now not deeply
studied - relative calibration between ROSAT and Beppo-SAX instruments.
Moreover, it allows the use of an extended energy range of both instruments,
where the response matrices are currently not well calibrated yet, but
where a significant amount of counts still exists.
A PSPC 3C273 spectrum has been extracted
from a pointing observation included in the public archive.
The spectral ratios are shown in
Fig.~\ref{fig6}. MECS spectrum remains pretty flat till $\simeq 1$~keV,
being inconsistent with the PSPC one also in the extended energy range,
where the PSPC flux is $\sim 3$ times higher.
Such a difference cannot
be explained in terms of spectral pivoting; in such a case the pivot should
lie at $E \simeq 5.4$~keV, while we found no evidence of
spectral steepening in the Beppo-SAX data.

In Fig.~\ref{fig7} the X--ray SED of the PSPC (0.1--2.4~keV) and
\begin{figure}
\epsfig{figure=H0849.f7,height=6.0cm,width=9.0cm,angle=0}
\caption{X--ray SED ($\nu F_{\nu}$ in units of erg~cm$^{-2}$~s$^{-1}$~Hz$^{-1}$)
of the ROSAT/PSPC, Beppo-SAX/MECS and 1996 July ASCA/SIS0
observations}
\label{fig7}
\end{figure}
MECS (1,5--10.5~keV) data is plotted, together with one
ASCA (0.5--10~keV) observation performed during summer 1996
(a complete analysis of the ASCA data is included in
Turner et al. 1998).
ASCA spectrum is consistent with
and rather flat ($\Gamma = 1.4-1.6$) power-law in
the whole 0.7--10~keV band and a soft excess at lower
energies, although with a much
lower flux than the PSPC spectrum (Turner et al. 1998).
The 0.7--2~keV spectrum had undergone a transition from
a steep ($\Gamma \simgt 2.5$) to a flat ($\Gamma \simeq
1.55$) state.
Two possible explanations are viable to account for the
observed behavior. 1H0419-577 could display a remarkably
variable soft excess.
Alternatively, ROSAT and Beppo-SAX
could have monitored two different states of the nuclear
primary continuum source. 

\section{Discussion}

1H0419-577 displays a combination of X--ray continuum
and line properties which is somewhat at odds with other
Seyfert 1s. A simple power--law is a good description of the data in the
whole 1.8--40~keV band, with a rather flat index ($\Gamma = 1.55$).
Both thermal and non--thermal Comptonization models can produce
such flat spectra
(Svensson 1994, Haardt et al. 1997), although
in rather extreme conditions.
No iron emission line was detected: the upper limit on the EW
of a narrow (broad) iron line from neutral iron is $\sim$90 (250) eV.
We extrapolated from the ASCA Nandra et al. (1997a) sample
the 5 objects whose 3--10~keV photon index is consistent with 1H0419-577
one within the statistical uncertainties
(cfr. Table 2 in their paper: NGC3227, NGC3783, Mrk841, NGC6814 and
MCG-2-58-22). The average observed EW in these objects
($121\pm16$) is significantly
higher then the 1H0419-577 upper limit. The constraints on the line
properties are however not very tight. It is worthwhile to notice
that the line upper limits are consistent, within the statistical
uncertainties, both with the EW vs. luminosity anti-correlation discovered
by Nandra et al. (1997b) and with the line detected in the ASCA spectrum
by Turner et al. (1998). 

The 1H0419-577 2--10~keV luminosity is
$\sim 5 \times 10^{44}$ erg s$^{-1}$, which points to a borderline
case between Seyferts and quasars. The lack of Compton reflection
and/or fluorescent iron line in the X--ray spectra of quasars
is still puzzling, given the fact that these features seem almost
universal in low luminosity objects. A possible explanation assumes that
the disk surrounding the central black hole becomes substantially
ionized when the primary luminosity increases
(Matt et al. 1993). That would decrease the contrast
between reflected and direct continua and shift the line
emission centroid towards energies corresponding to
ionized iron stages; eventually, the intensity of the
iron line would fade away when iron turns more and more
into a completely ionized
stage. No ionized line or Compton reflection are required by
the 1H0419-577 data, but the constraints on the spectral parameters
are again too loose.

The comparison between MECS results and
1992 ROSAT and 1996 ASCA observations demonstrates
that {\it at least} the 0.5--2.5~keV has changed
by a factor of $\sim 6$ in flux while turning from a
soft ($\Gamma_{PSPC} \simgt 2.5$) to a hard ($\Gamma_{MECS} \simeq 1.55$)
state. ASCA data suggested the presence of a soft excess above
the extrapolation of a rather hard high--energy power--law below
0.7~keV (Turner et al. 1998),
which would have had a much lower flux and/or effective temperature
than observed by ROSAT.
Some authors (Walter \& Fink 1993, Puchnarewicz et al. 1996)
suggested that the optical to soft X--ray
continuum could be seen as part of the same ``Big Bump'' (BgB), to
which a hard power--law with typical $\Gamma = 2$ is underlying.
Variability by a factor of $\sim$few seems to be a common
property of soft X--ray radio--quiet AGN (Mannheim et al. 1996)
It was recently suggested that such a variability is connected
to the transient nature of disk emission around $10^6$-$10^7 \ M_{\odot}$
black holes with near-Eddington accretion (Grupe et al. 1998).
In order to qualitatively check these
scenarios, we show the optical to X--ray SED in Fig.~\ref{fig10}.
\begin{figure}
\epsfig{figure=H0849.f8,height=6.0cm,width=9.0cm,angle=0}
\caption{Optical to X--ray SED. Both Beppo-SAX MECS ({\it filled
circles}) and ROSAT/PSPC ({\it empty squares}) data are shown together with
the Beppo-SAX simultaneous optical spectrum}
\label{fig10}
\end{figure}
Simultaneous MECS and optical data seem to suggest that the peak of the
energy density lies around $\sim 3500-4000$~\AA, as typically observed
in Radio Quiet Quasars (RQQ, Laor et al. 1997). If the PSPC data
represents a soft excess which is still present also
at the epoch of the
Beppo-SAX observation, a simple extrapolation of the data suggests
that such a soft excess 
cannot be reconnected in a single component with the
optical to hard X--ray spectrum.
The soft X--rays are clearly variable, and a
corresponding dynamics of the optical/UV emission are required to
occur as well if the optical emission is tightly related to it.
However, strong variations in the optical band
have not been observed yet.

A way to characterize the soft X--ray and optical properties of
QSO is through the multiwavelength ``point-to-point'' spectral indices
$\alpha_{ox}$ and $\alpha_{os}$. We follow hereinafter the definition by
Laor et al. (1997), where
$\alpha_{os} = [\log(f_{0.3 \ keV}/f_{3000 \AA})]/1.861$
and $\alpha_{ox} = [\log(f_{2 \ keV}/f_{3000 \AA})]/2.685$
($f$ is the flux density).
We extrapolated the density flux at 3000\AA from the
optical spectrum presented in Sect.~2.
For 1H0419-577 $\alpha_{os} = -0.93$
and $\alpha_{ox} = -1.23$. These values are in good agreement
with the outcomes of the RIXOS sample analysis (Puchnarewicz
et al. 1996), but flatter than in the complete PG sample of
Wilkes et al. (1994) and in the optically selected sample of
Laor et al. (1997) (for both $\alpha_{ox} \simeq -1.5$).
A possible explanation for this discrepancy is that optically
selected samples tend to be biased in favor of objects with
intense optical/X--ray ratios. 1H0419-577 $\alpha_{ox}$ is also
consistent, within the statistical dispersion, with Walter \& Fink sample
(1993, $\alpha_{ox} = -1.37 \pm 0.23$), which is selected upon the X--ray
brightness (the difference in $\alpha_{ox}$ induced by the different choice
of the reference optical energy between Laor et al. (1997) and
Walter \& Fink (1993) - 3000\AA vs. 2650\AA - is well within the statistical
dispersion).

The four year variation
timescale implies that the soft excess emission region is
smaller than 1 pc, and therefore associated with the nuclear region,
where the primary continuum is expected to undergo strong
reprocessing. If reprocessing is the ultimate cause of
the soft excess,
the energy balance requires that any cut-off
is confined at energies $\simgt$100~keV; that
does not contrast the average behavior of
Seyfert galaxies (Gondek et al. 1996).
Any change of the flux/temperature of the
soft excess has to be causally related to the variation of the high-energy
spectrum.
The lack of response of the soft X--rays to
a $\Gamma \sim 0.2$ change of the 0.7--10~keV
photon index within two weeks (Turner et al. 1998)
puts any reprocesser farther than 10$^4$ Schwarzschild radii
from a 10$^7 \ M_{\odot}$ black hole.
Domination of scattering
in the soft X--ray emission
is unlikely because of the variability in the PSPC
and HRI data (Turner et al. 1998).

An appealing alternative possibility is that the
difference between the measured steepness of the
PSPC and MECS spectra ($\Delta \Gamma > 0.9$) is due
to a change of the primary continuum itself.
Haardt \& Maraschi (1993) have shown that a disk corona,
Comptonizing the soft photons supplied by a Shakura-Sunayev
disk, can naturally reproduce the medium X--ray spectra shape typically
observed in Seyfert 1.
Interestingly enough, $\Gamma_{hard}$ and $\Gamma_{soft}$
in 1H0419-577 are very close
to the expected spectral indices if the coronal plasma undergoes
a transition between a scattering optically thin ($\tau \simlt 1$, $\Gamma = 1.55$)
and thick ($\tau \sim 1$, $\Gamma = 2.5$) regime (Haardt, Maraschi \& Ghisellini, 1997).
A pair dominated corona would
require an increase of the 2--10~keV flux by more than 4 orders of
magnitude for a steepening $\Delta \Gamma \sim 1$ to be achieved.
On the contrary,
if we extrapolate the PSPC best fit power-law into
the 2--10~keV band,
the luminosity is $\sim 4.8 \times 10^{44}$~erg~s$^{-1}$, which is comparable
with the MECS measurement.
A pair-saturated plasma can be therefore ruled out.
A measurable side-effect in such a scenario would
be an increase by an order of magnitude of the ratio between the
flux in the 2--10~keV band above $E \simgt 30$
keV.
The hypothesis of a change of the Comptonized spectral shape could provide
also a good explanation for the relative faintness of the iron line emission,
if the Compton component responds not simultaneously
to the - basically unknown - pattern
of variability of the primary emission.

Another possibility to explain a change in the primary continuum
spectral shape comes from the analogy between Seyfert galaxies and
Black Hole Candidates (BHC).
BHC are well know to display two different intensity and spectral
states: a high and soft one,
characterized above $\sim$ 10~keV by a power-law spectrum with photon
index $\simeq 2.5$, and a low and hard one, with typical $\Gamma
\sim 1.3-1.9$. 
Such an analogy had been first
suggested to explain the very soft ($\Gamma \simeq 2.6$) ASCA spectrum
of the NLSy1 galaxy RE1034+39 (Pounds et al. 1995). 

Ebisawa et al. (1996) have
recently suggested that the difference between the two states is due
to different Comptonization mechanisms
in a viscuosless shock two--phase accretion disc. Thermal
Comptonization in a disk with a very low accretion rate (and
therefore low thermal soft emission) would yield a
hard spectrum with $\Gamma \sim 1.5$,
regardless of the {\it absolute}
black hole mass. If, however, the accretion rate increases, more
soft photons are supplied,
and the post-shock region cooling becomes more efficient.
The cooler inflow towards the nuclear black hole is responsible for
a steep non--thermal emission.
The resulting spectrum in a quasi-Eddington regime is the combination of the
thermal emission from the optically thick disk and of a power-law with
$\Gamma \sim 2.5$. In this framework, the observed variability pattern
in 1H0419-577 would therefore suggest a transition
phase from a bulk motion to a thermal motion
regime due to a change of the accretion rate.
Typical $kT_{BB}$ in BHC in high state are $\sim 1$~keV;
scaling $T_{B} \propto M^{1/4}$, a blackbody temperature
$\simlt 80 \ eV$ corresponds to a black body mass $M_{BH} \simgt 2 \times
10^6 \ M_{\odot}$. The bolometric luminosity
of the blackbody component as measured by PSPC on 1H0419-577 is
$\sim 5 \times
10^{44} \ erg \ s^{-1}$ and self-consistently in this case
$\dot{m}/\dot{m}_{Edd} \sim 1$.
However, it must be kept in mind that the determination of
a blackbody temperature sensitively depends on the energy band
where the spectral fits are performed and that the spectral
deconvolution in the PSPC spectra of 1H0419-577 is not unique.
We are then far from considering this coincidence as a confirmation of
the proposed scenario.

\section{Summary}

We report on the first simultaneous
X--ray and optical observations of the
radio--quiet AGN 1H0419-577. The main results can be summarized as follows:

\noindent
(1) the optical spectrum points unambiguously to a classification
as a Seyfert 1.

\noindent
(2) 1H0419-577 is detected in X--rays up to 40~keV. The whole 1.8--40~keV
spectrum is well modeled by a simple power--law with a rather flat
($\Gamma \simeq 1.55$) spectral index.

\noindent
(3) No iron emission line or Compton reflection
by an X--ray illuminated neutral disc
are required by the data.
However, the constraints on the spectral parameters
are not so tight to rule out the possibility that
this object follows the trend of high luminosity
Seyferts to have fainter and/or ionized iron lines and
Compton continua.

\noindent
(4) The SED of the simultaneous optical and X--rays observations
peaks apparently around 3500-4000\AA, as typically observed in RQQ
(Laor et al. 1997).

\noindent
(5) The comparison with a 1992 ROSAT and a pair of 1996
ASCA observations (Turner et al. 1998) reveals
that the spectrum has undergone a transition
between a soft ($\Gamma \simgt 2.5$) and a hard
($\Gamma \simeq 1.55$) state, {\it at least} in the
0.7--2~keV band. We have discussed in this paper
two possible explanations for this behavior
and their implications:
\begin{itemize}
\item[-] a remarkably variable
($\sim$ half an order of magnitude in flux and/or
effective temperature) soft X--ray excess, which should come along
with a corresponding dynamics of the optical and UV emission,
if the SED is dominated by a single BgB. The
optical historical monitoring of this source, albeit sparse,
seems to rule out this possibility;
\item[-] a change of the primary continuum spectrum,
which could be explained either as a transition from
a Compton thick to a Compton thin regime in the disk-corona
two-phase model of Haardt \& Maraschi (1993) or a
transition between a bulk to
a thermal motion regime in a viscuosless two--phase
shock accretion disc (Chakrabarti and Titarchuk 1995;
Ebisawa et al. 1996).
\end{itemize}

\begin{acknowledgements}

The authors would like to warmly thank the whole Beppo-SAX team, whose
intelligent and hard work made possible this observation to be 
successfully performed. The authors acknowledge valuable suggestions
from M.Cappi, P.Giommi and F.Haardt
and a careful revision of the manuscript by S.Kaiser.
M.G. acknowledges an ESA Research Fellowship.
A.C. and G.M.S.
acknowledge financial support from
ASI contract ARS96-70. This work has made use of
the NASA/IPAC Extragalactic Database (NED),
which is operated by the Jet Propulsion Laboratory, Caltech, under
contract with N.A.S.A., and of the
cleaned and linearized event files produced at the Beppo-SAX
Science Data Center.

\end{acknowledgements}

\end{document}